\documentclass[%
 reprint,
 amsmath,amssymb,
 aps,prl
]{revtex4-2}

\usepackage{graphicx}
\usepackage{dcolumn}
\usepackage{bm}
\usepackage{simplewick}
\usepackage[dvipsnames]{xcolor}
\usepackage{braket,cancel}
\definecolor{lgray}{gray}{0.4}
\definecolor{cor}{rgb}{0.94, 0.41, 0.35}
\definecolor{mag}{rgb}{0.7451, 0.2039, 0.3333}
\usepackage[colorlinks=true,urlcolor=mag,linkcolor=mag,
citecolor=mag,pdfpagelabels=true,hypertexnames=true,
plainpages=false,naturalnames=false]{hyperref}
\newcommand{\x}{{\boldsymbol x}}

\def\k{{\boldsymbol k}}

\def\bea{\begin{eqnarray}}
\def\eea{\end{eqnarray}}
\def\be{\begin{equation}}
\def\ee{\end{equation}}
\def\ba{\begin{array}}
\def\ea{\end{array}}
\def\nn{\nonumber}

\newcommand{\de}{{\rm d}}

\begin{document}

\preprint{APS/123-QED}

\title{Revisiting stochastic inflation with perturbation theory}

\author{Gonzalo A. Palma$^a$}

\author{Spyros Sypsas$^{b,c}$}

\affiliation{
$^a${\it Departamento de F\'isica, FCFM, Universidad de Chile, Blanco Encalada 2008, Santiago, Chile}
\\
$^b${\it High Energy Physics Research Unit, Faculty of Science, Chulalongkorn University, Bangkok 10330, Thailand}
\\
$^c${\it National Astronomical Research Institute of Thailand, Don Kaeo, Mae Rim, Chiang Mai 50180, Thailand}
}

\begin{abstract}  
A long-standing problem in primordial cosmology is to understand the precise relation between the stochastic formalism and standard perturbation theory for light scalar fields in inflationary spacetimes. A complete correspondence between the two frameworks has remained elusive, even for a single self-interacting spectator field on a fixed de Sitter background. In this Letter, we revisit the assumptions underlying the derivation of the Langevin and Fokker--Planck equations that form the basis of the stochastic approach. We show that the standard stochastic treatment is effectively equivalent to neglecting super-long-wavelength modes beyond the observable range described by these equations. Using perturbation theory, we quantify how these modes enter through virtual loop effects and demonstrate that, once treated consistently, they induce definite corrections to both the Langevin and Fokker--Planck equations.

\end{abstract}

                          
\maketitle

\noindent \emph{\bf Introduction.--} Correlation functions of a light scalar field in inflationary backgrounds, computed within standard perturbation theory (SPT), exhibit time-dependent secular growth in the form of powers of $\ln a(t)$, where $a(t)$ is the scale factor~\cite{Linde:1982uu,Starobinsky:1982ee,Vilenkin:1982wt,Ford:1984hs,Antoniadis:1985pj,Weinberg:2005vy,Weinberg:2006ac,Burgess:2009bs,Seery:2010kh,Polyakov:2012uc,Hu:2018nxy,Kitamoto:2018dek}. Their presence signals the breakdown of perturbation theory on length scales far beyond the horizon. Predicting the statistics of the field on these scales therefore requires resuming contributions from diagrams at all orders, a practically unattainable task. A widely adopted strategy to address this difficulty is provided by the stochastic formalism, originally introduced by Starobinsky and Yokoyama~\cite{Starobinsky:1986fx,Starobinsky:1994bd}. In this framework, the superhorizon dynamics of $\varphi$ is described by an effectively classical field driven by a stochastic noise that accounts for the influence of short-wavelength quantum fluctuations. This description leads to a Fokker--Planck equation governing the probability density function (PDF) of a suitably coarse-grained field.

The stochastic formalism and SPT are expected to coincide within a window of scales where quantum fluctuations are still relevant, while nonlinear interactions remain under perturbative control. However, despite extensive work, the precise manner in which these two descriptions overlap remains only partially understood. Although their equivalence has been established in specific limits and under particular assumptions~\cite{Onemli:2002hr,Miao:2005am,Baumgart:2019clc, Palma:2023idj,Tsamis:2005hd, Guilleux:2015pma, Finelli:2008zg, Riotto:2011sf, PerreaultLevasseur:2013kfq, Burgess:2014eoa, Burgess:2015ajz, Moss:2016uix, Garbrecht:2013coa, Garbrecht:2014dca, Tokuda:2017fdh, Gorbenko:2019rza, Mirbabayi:2019qtx, Collins:2017haz, Vennin:2015hra, Honda:2023unh, Cohen:2020php, Mirbabayi:2020vyt, Baumgart:2019clc, Cohen:2021fzf, Green:2022ovz, Cohen:2022clv}, a fully systematic correspondence is still lacking.

In this Letter, we take a step toward clarifying this connection. As we shall see, the derivation of the Fokker--Planck equation requires a decomposition of the scalar field of the form
\be
\label{splitting-us}
\varphi = \varphi_S + \varphi_L + \varphi_{\rm IR}.
\ee
$\varphi_S$ contains modes with wavelengths shorter than the coarse-graining length that sets the resolution of the system. This scale is taken to be a fixed physical length, slightly larger than the Hubble radius. $\varphi_L$ includes observable degrees of freedom whose statistics are described by the Fokker--Planck equation; in practice, its maximum wavelength is often specified by introducing a comoving cutoff $k_*$ in momentum space. Finally, $\varphi_{\rm IR}$ denotes the collection of super-long-wavelength modes that are inaccessible to local observers.
As we shall demonstrate, the standard stochastic formalism~\cite{Starobinsky:1986fx}, when applied to the statistics of $\varphi_L$, effectively corresponds to setting $\varphi_{\rm IR}=0$. Our central claim is that this procedure is incomplete. The field $\varphi_{\rm IR}$ influences the statistics of $\varphi_L$ through nonlinear interactions, and its effects cannot, in general, be neglected. 

From the perspective of SPT, setting $\varphi_{\text{IR}} = 0$ amounts to discarding loop contributions with momenta below the infrared (IR) comoving scale $k_\ast$. However, $k_\ast$ --which delineates the range of external momenta whose statistics is probed-- cannot simultaneously serve as an infrared cutoff for virtual processes such as loop corrections, unless the degrees of freedom in $\varphi_{\text{IR}}$ are properly integrated out. In other words, it cannot act as an IR cutoff for both external and internal momenta. 

Below, we examine how these infrared modes affect the stochastic description of $\varphi_L$. We show that, once $\varphi_{\rm IR}$ is treated consistently, the resulting Fokker--Planck equation acquires corrections, which are particularly relevant in the nonlinear regime~\cite{Ezquiaga:2019ftu, Vennin:2020kng, Hooshangi:2021ubn, Achucarro:2021pdh, Cai:2022erk, Pi:2022ysn,Hooshangi:2023kss}. More broadly, our conclusions align with recent work~\cite{Cruces:2018cvq, Cruces:2021iwq, Cruces:2022imf,Figueroa:2021zah,Launay:2024qsm}
highlighting that several foundational assumptions of the stochastic
formalism must be revisited when nonlinearities and mode couplings are
treated with full consistency.


\vspace{3pt}
\noindent\emph{\bf Stochastic formalism.--}
We begin by reviewing the stochastic approach used to determine the statistics of scalar fluctuations in quasi-de Sitter spacetimes. We work with a flat FLRW metric 
\be
\de s^2 = -\de t^2 + a^2(t)\de{\x}^2,
\ee
where $a(t)$ is the scale factor. The Hubble expansion rate is $H = \dot a / a$. The field theory we consider is
\be
\label{action}
S = \int \! \de^3 x \, \de t \,  \frac{a^3}{2} 
\left[
\dot \varphi^2
- \frac{1}{a^2}(\nabla \varphi)^2
- 2 \mathcal V(\varphi, t) + \cdots
\right] ,
\ee
where the potential $\mathcal V(\varphi,t)$ encodes nonlinear interactions and may be expanded as
\be
\label{Taylor-V}
\mathcal V(\varphi,t)
=
\sum_{n=3}^{\infty}
\frac{\lambda_n(t)}{n!}\,\varphi^n.
\ee
The ellipses in (\ref{action}) represent gradient operators due to gravitational interactions that are commonly disregarded. The equation of motion derived from \eqref{action} is
\be
\label{eq-of-motion}
\ddot \varphi
+  3H  \dot \varphi
- a^{-2}\nabla^2\varphi
+ \mathcal V'(\varphi)
=0 ,
\ee
where the prime denotes a partial derivative with respect to $\varphi$. To parameterize deviations from exact de Sitter spacetime, for which $H$ is constant, we employ the standard slow-roll parameters
\be
\epsilon \equiv -\frac{\dot H }{ H^2}, \qquad \eta \equiv \frac{\dot \epsilon}{H \epsilon} .
\ee
We will assume $\epsilon \ll 1$ and $|\eta| \ll 1$. In general, the couplings $\lambda_n$ may depend on $\epsilon$, $\eta$, and higher-order slow-roll parameters. We will not need the precise relation between these coefficients and the slow-roll hierarchy; instead, we simply assume that they vary slowly $| \dot \lambda_n / H\lambda_n | \ll 1$,
as is typical during slow-roll inflation.

To continue, we split the field $\varphi$ as in Eq.~\eqref{splitting-us}. To this end, it is convenient to write the coarse-grained field as
\be \label{f_L}
\varphi_L({\x},t)=\int_{\k} e^{ i {\k} \cdot {\x}}\, W(k , t)\, \tilde \varphi_{\k} (t),
\ee
where $\int_{\k}\equiv\int\frac{\de^3 k}{(2\pi)^3}$. Here, $\tilde \varphi_{\k}(t)$ is the Fourier transform of $\varphi({\x},t)$ and $W(k,t)$ is a window function specifying the splitting. We choose~\cite{Tsamis:2005hd}
\be
\label{window-2}
W (k,t) = \theta\!\left(k_*(t)-k\right)\,\theta\!\left(k-k_*(t_{\rm i})\right),
\ee
with $\theta(x)$ the Heaviside step function~\footnote{More rigorously, this split should be performed with a smooth window function. See for instance~\cite{Gorbenko:2019rza} for a detailed discussion on this point.} and $k_*(t)=\sigma H(t) a(t)$, where $\sigma>0$ is a small dimensionless parameter. The scale $k_*(t_{\rm i})$ corresponds to the comoving infrared cutoff $k_*$ mentioned in the Introduction and is chosen such that the support of $\varphi_L$ is empty at the initial time $t=t_{\rm i}$. As a result, $\varphi_L$ contains only modes in the range $k_*(t_{\rm i})<k<k_*(t)$, while $\varphi_S$ and $\varphi_{\rm IR}$ account for the rest.

The stochastic framework relies on the idea that short-wavelength modes may be treated as linear quantum fluctuations, whereas long-wavelength modes behave classically but nonlinearly. To implement this separation, three key assumptions are made~\cite{Starobinsky:1986fx}. The first is that $\varphi_{\rm IR}=0$, which amounts to the observable long-wavelength field $\varphi_L$ being insensitive to modes of even longer wavelength. The second is that $\ddot \varphi_L$ can be neglected against $3 H \dot \varphi_L$. Finally, the third assumption is that for wavelengths shorter than $2\pi/k_*(t)$, $\varphi_L$ acts as a background field. In other words Fourier modes of $\varphi_S$ respect:
\begin{equation}
\label{eq-of-motion_short}
\ddot {\tilde\varphi}_{\k} + 3 H \dot {\tilde\varphi}_{\k}
+ a^{-2} k^2 \tilde\varphi_{\k}
+ \mathcal V''(\varphi_L)\tilde\varphi_{\k}=0 .
\end{equation}
The general solution to this equation, consistent with canonical commutation relations, is $\tilde \varphi_{\k}(t)
= f(k,t)\,\hat a_{\k}+f^*(k,t)\,\hat a^\dag_{-{\k}}$,
where $[\hat a_{\k},\hat a^\dag_{{\k}'}]=(2\pi)^3\delta({\k}-{\k}')$, and with the mode function $f(k,t)$ fixed by the Bunch--Davies initial condition. 

Implementing these assumptions, Eq.~\eqref{eq-of-motion} turns into the following Langevin equation
\be
\label{L-eq}
\dot \varphi_L + \frac{1}{3H}\,\mathcal V'(\varphi_L)
= H\,\widehat\xi(t),
\ee
where $\widehat\xi(t)$ is a Gaussian stochastic noise defined as $\widehat\xi(t)
\equiv H^{-1}\int_{\k} \dot W(k,t) \tilde\varphi_{\k}(t)$. By further noticing that $\dot W(k,t)=H (1 - \epsilon) k_*(t)\,\delta(k_*(t)-k)$, the two-point correlation of the noise is found to be
\be \label{norm-xi}
\left\langle \widehat\xi(t)\widehat\xi(t')\right\rangle
=\frac{H (1-\epsilon)}{4\pi^2}\, \left( \frac{2 k^3|f|^2}{H^2} \right)_{\! k = k_*(t)} \!\!\!\!\!\!\!\!\!\! \delta(t-t').
\ee 
Equations (\ref{L-eq}) and (\ref{norm-xi}) form the backbone of the stochastic description of $\varphi_L$. 

Relaxing the second and third assumptions leads to relevant corrections to the stochastic dynamics~\cite{Cruces:2018cvq,Cruces:2022imf,Figueroa:2021zah,Cruces:2021iwq,Launay:2024qsm} that are orthogonal to the ones studied here. The main purpose of this Letter is to question the validity of the first assumption, that is $\varphi_{\rm IR}=0$. Before doing so, let us rederive a well-known result within the framework of stochastic inflation: the Fokker--Planck equation.

\vspace{3pt}
\noindent\emph{\bf PDFs in the stochastic formalism.--} Equations~(\ref{L-eq}) and~(\ref{norm-xi}) provide the starting point for averaging over realizations of the stochastic noise and translating the stochastic dynamics into a deterministic evolution equation for the PDF describing the single-point statistics of $\varphi_L$ (see, for instance, Ref.~\cite{Pinol:2018euk}). The result is a Fokker--Planck equation. Here we derive it following a different strategy based on perturbation theory. 

In the absence of $\mathcal V$, the field $\varphi_L$ is trivially Gaussian, with a distribution given by
\be
\label{Gauss}
\rho_0(\varphi,t)
=\frac{1}{\sqrt{2\pi\sigma_L^2(t)}}
\exp\!\left[-\frac{\varphi^2}{2\sigma_L^2(t)}\right] .
\ee
The variance, which is $\sigma_L^2(t)=\left\langle \varphi_L(t)\varphi_L(t)\right\rangle$, may be computed by first writing $
\varphi_L(t)=\int_{t_i}^{t} \de t'\,H (t')\widehat\xi(t')$ and then employing Eq.~(\ref{norm-xi}). Performing the change of variables $t'\to k(t')=\sigma H(t')a(t')$, the time integral extends over the interval $[t_{\rm i},t]$, leading to the appearance of the window function~\eqref{window-2}, and hence
\be \label{var-f}
\sigma_L^2(t)
= \frac{1}{2\pi^2}
\int_0^\infty \frac{\de k}{k}\, k^3 |f(k,t)|^{2} W(k,t).
\ee 
In the particular case of de Sitter, where $\epsilon = 0$ (and therefore $H$ is a constant), one has $
f(k,t)  = \frac{H}{\sqrt{2 k^3}} (1 + i k/aH) e^{- i k/aH}$, and so $k^3 |f(k,t)|^{2}=H^2/2$ on the range $k_*(t_{\rm i})<k<k_*(t)$. As a result, the variance is explicitly given by
\be
\label{var}
\sigma_L^2(t)
=\frac{H^2}{4\pi^2}
\ln\frac{k_*(t)}{k_*(t_{\rm i})}
=\frac{H^3}{4\pi^2}(t-t_{\rm i}) .
\ee
Consequently, at $t=t_{\rm i}$ the PDF reduces to a Dirac delta function, $\delta(\varphi)$, and subsequently diffuses in field space. In other words, as time progresses, more and more short-wavelength modes enter the observable domain defining $\varphi_L$, thereby increasing the uncertainty in its amplitude at a given spatial point. This conclusion persists in the case of quasi-de Sitter, with small corrections.

Let us now consider the case $\mathcal V \neq 0$. Integrating~\eqref{L-eq} and iterating it once, we obtain
\be
\label{stoch-phi-L}
\varphi_L(t)
= \varphi_G(t)
- \frac{1}{3}\int_{t_{\rm i}}^{t}\! \frac{\de t'}{H(t')}\,
{ \mathcal V}'
\!\left[\varphi_G(t')\right] + \mathcal O \left( \mathcal V^2 \right),
\ee
where $\varphi_G(t)=\int_{t_{\rm i}}^{t}\de t'\, H (t'´)  \widehat\xi(t')$ is a Gaussian field with a variance given by~\eqref{var-f}. Upon Taylor-expanding $\mathcal V$ as in (\ref{Taylor-V}), equation~\eqref{stoch-phi-L} allows us to compute the $n$-th cumulant
$\left\langle\varphi_L^n(t)\right\rangle_c$. Excluding the free Gaussian part, the cumulants follow from performing all possible Wick contractions
\(
\bcontraction[.5ex]{}{\varphi_G}{(t)}{\varphi_G}
\varphi_G(t)\varphi_G(t')
\)
and
\(
\bcontraction[.5ex]{}{\varphi_G}{(t')}{\varphi_G}
\varphi_G(t')\varphi_G(t')
\),
leading to
\bea
\label{eq:from-langevin}
\Big\langle \varphi_L^n(t) \Big\rangle_c
&=&
- \frac{n}{3}
\int_{t_{\rm i}}^{t}\!\frac{\de t'}{H(t')}
\Big(\langle\varphi_G(t)\varphi_G(t')\rangle\Big)^{n-1}
\nn\\
&&\times
\sum_{\ell=0}^{\infty}
\frac{\lambda_{n+2\ell} (t')}{\ell!}
\left(
\frac{1}{2}
\langle\varphi_G(t')\varphi_G(t')\rangle
\right)^{\ell}.\qquad
\eea
Diagrammatically, the contractions
\(
\bcontraction[.5ex]{}{\varphi_G}{(t')}{\varphi_G}
\varphi_G(t')\varphi_G(t')
\)
correspond to single-vertex loops with $\ell$ denoting their number. From the property (\ref{norm-xi}) respected by the noise $\widehat\xi(t)$, it follows that, as long as $t' \leq t$,
\be \label{variances-staro}
\langle\varphi_G(t')\varphi_G(t')\rangle
=\langle\varphi_G(t)\varphi_G(t')\rangle
=\sigma_L^2(t') .
\ee

To obtain an explicit expression for the cumulants, one needs to insert
\eqref{variances-staro} into \eqref{eq:from-langevin} and integrate over
time. To isolate the leading logarithmic time dependence of these cumulants,
it is useful to focus on the simpler de Sitter case, where $\sigma_L^2$
reduces to \eqref{var}. In this limit, the cumulants become
\be
\label{stoch_cum}
\!\!\Big\langle \varphi_L^n(t) \Big\rangle_c
=
-\frac{4\pi^2 n}{3H^4}
\,\sigma_L^{2n}(t)
\sum_{\ell=0}^{\infty}
\frac{\lambda_{n+2\ell}}{(n+\ell)\,\ell!}
\left(\frac{\sigma_L^2(t)}{2}\right)^{\ell}
\ee
This shows that the $\ell$-loop contribution to
$\langle\varphi_L^n(t)\rangle_c$ is infrared-enhanced by a factor proportional
to $[\ln a(t)]^{n+\ell}$.

The cumulants in \eqref{stoch_cum} determine the PDF
$\rho(\varphi,t)$ through the Edgeworth expansion
\be
\label{pdf-qft}
\!\rho(\varphi ,t)
=
\rho_0(\varphi ,t)
\left[
1+
\sum_{n=1}^{\infty}
\frac{\langle\varphi_L^n(t)\rangle_c}{n!\,\sigma_L^{n}}
\,{\rm He}_n\left(\frac{\varphi}{\sigma_L}\right)
\right],
\ee
where $\rho_0$ is the Gaussian function \eqref{Gauss}, and ${\rm He}_n(x)$
denotes the probabilistic Hermite polynomial of order $n$. Substituting
\eqref{stoch_cum} into \eqref{pdf-qft} and taking a time derivative of the
resulting distribution, one obtains the desired Fokker--Planck equation
\begin{eqnarray}
\label{Fokker-Planck}
\dot\rho
= \frac{H^3}{8\pi^2}\rho'' + \frac{1}{3H}
\!\left(\rho\, \mathcal{V}' \right)'.
\end{eqnarray}
This coincides with the well-known result of Starobinsky~\cite{Starobinsky:1986fx},
describing the time evolution of the distribution $\rho$ under the influence
of the potential $\mathcal V$~\footnote{An alternative way to establish the
relation between Eqs.~\eqref{stoch_cum} and~\eqref{Fokker-Planck} is to
multiply both sides of Eq.~\eqref{Fokker-Planck} by
${\rm He}_n(\varphi/\sigma_L)$ and integrate over $\varphi$. This gives
$\frac{\de}{\de t}\langle \varphi^n \rangle_c
=
n(n-1)\frac{H^3}{8\pi^2}\langle \varphi^{n-2}\rangle_c
-\frac{n}{3H}\sum_{m=2}^{\infty}\frac{\lambda_m}{(m-1)!}
\langle \varphi^{m+n-2}\rangle_c$, which provides a nonlinear hierarchy
relating the time evolution of all cumulants. Solving this hierarchy
reproduces Eq.~\eqref{stoch_cum}.}.


\vspace{3pt}
\noindent\emph{\bf Restoring infrared modes.--} Now comes the main point of our Letter. In deriving the Langevin equation~\eqref{L-eq} from the full quantum equation of motion~\eqref{eq-of-motion}, we have implicitly replaced $\mathcal V'(\varphi)$ by $\mathcal V'(\varphi_L)$, thereby neglecting the dependence on the infrared field $\varphi_{\rm IR}$. As a consequence, the contractions entering~\eqref{eq:from-langevin} involve only the restricted Gaussian field $\varphi_G$. What changes once the contribution of $\varphi_{\rm IR}$ is taken into account? 

To address this question, it is useful to introduce the filtering operator
\be
\widehat W\!\left\{\mathcal F(\x,t)\right\}
\equiv
\int_{\k} e^{i\k\cdot\x}\,W(k,t)\,\tilde{\mathcal F}_{\k}(t),
\ee
where $W(k,t)$ is the window function defined in~\eqref{window-2} and $\tilde{\mathcal F}_{\k}(t)$ denotes the Fourier transform of $\mathcal F(\x,t)$. In this notation, the observable field is simply $\varphi_L(\x,t)=\widehat W\{\varphi(\x,t)\}$. Applying $\widehat W$ to the full equation of motion~\eqref{eq-of-motion}, assuming again that the short modes evolve linearly, and neglecting second time derivatives of $\varphi_L$, one finds
\be
\label{L-eq-IR}
\dot\varphi_L
+\frac{1}{3H}\,\widehat W\!\left\{\mathcal V'(\varphi)\right\}
=H \,\widehat\xi(t),
\ee
where $\varphi$ is understood as the full field $\varphi=\varphi_S+\varphi_L+\varphi_{\rm IR}$. At this point, a crucial difference with respect to the standard stochastic equation~\eqref{L-eq} becomes apparent: in general,
\be
\widehat W\!\Big\{\mathcal V'(\varphi)\Big\}
\neq
\mathcal V'\!\Big(\widehat W\{\varphi\}\Big).
\ee
Therefore, the standard stochastic formalism implicitly assumes that the coarse-graining operation $\widehat W\{\cdot\}$ commutes with the nonlinear potential $\mathcal V'(\cdot)$. This amounts to neglecting the interactions of the observable field $\varphi_L$ with both the short-wavelength sector $\varphi_S$~\cite{Cruces:2021iwq,Figueroa:2021zah} and the super-long-wavelength sector $\varphi_{\rm IR}$. Our focus is on the latter.

As in the previous section, we may compute the $n$-point cumulants of $\varphi_L$ to first order in $\mathcal V$. Integrating Eq.~\eqref{L-eq-IR} in time yields
\be
\label{QFT-varphi_L}
\varphi_L(t)
\simeq
\varphi_G(t)
-\frac{1}{3}\int_{t_{\rm i}}^{t}\!\frac{\de t'}{H(t')}
\widehat W\!\left\{\mathcal V'\!\left[\varphi(t')\right]\right\},
\ee
where, as before, $\varphi_G(t)=\int_{t_{\rm i}}^{t} \de t'\,H(t')\widehat\xi(t')$. The resulting cumulant then reads
\bea
\label{eq:from-qft}
\Big\langle \varphi_L^n(t) \Big\rangle_c
&=&
-\frac{n}{3}
\int_{t_{\rm i}}^{t}\! \frac{\de t'}{H(t')} \,
\Big(\langle\varphi_G(t)\varphi(t')\rangle\Big)^{n-1}
\nn\\
&&\times
\sum_{\ell=0}^{\infty}
\frac{\lambda_{n+2\ell} (t' )}{\ell!}
\left(
\frac{1}{2}\sigma_{\rm tot}^2 (t') 
\right)^{\ell},
\eea
where we have defined
\be
\label{deSitter-inv}
\sigma_{\rm tot}^2 (t') \equiv \langle\varphi(t')\varphi(t')\rangle .
\ee
This result can be alternatively obtained directly with SPT~\cite{Chen:2017ryl, Palma:2017lww, Chen:2018uul, Chen:2018brw, Palma:2025oux} by first computing $n$-point correlation functions $\langle\tilde\varphi_{\k_1}\cdots\tilde\varphi_{\k_n}(t)\rangle_c$, and then extracting cumulants by integrating in momentum space. Equation~\eqref{eq:from-qft} has the same structure as~\eqref{eq:from-langevin}, except that the field $\varphi(t')$ is no longer restricted to the range of scales contained in $\varphi_G$. Since the window function delimits only the external momenta accessible to observers, as long as $t>t'$, the mixed contraction is
\be
\label{mixed-contraction}
\langle\varphi_G(t)\varphi(t')\rangle=\sigma_L^2(t).
\ee
By contrast, the equal-time contraction $\sigma_{\rm tot}^2(t')=\langle\varphi(t')\varphi(t')\rangle$, which appears inside loops, is unrestricted. As a consequence, its time dependence must differ substantially from that of \eqref{mixed-contraction}. In fact, we claim that its time dependence can be at most slow-roll suppressed:
\be
\label{sigma-slow-roll-suppressed}
|\dot \sigma_{\rm tot} / H \sigma_{\rm tot}| \ll 1 .
\ee
Moreover, in the specific case of de Sitter, $\sigma_{\rm tot}^2$ must be a constant. In what follows we substantiate these claims.

\vspace{3pt}
\noindent\emph{\bf Loop regularization.--} There is little room to object to the structure of Eq.~\eqref{eq:from-qft}, which restores the effect of infrared modes in the computation of cumulants. A more delicate issue concerns our claim (\ref{sigma-slow-roll-suppressed}) related to internal loop contractions, in that $\sigma_{\rm tot}^2(t')$ must be taken to be almost constant. Let us therefore discuss this point in more detail.

To begin with, since we are interested in cumulants up to first order in $\mathcal V$, the unrestricted field operator $\varphi$ entering \eqref{eq:from-qft} may be evolved linearly. In Fourier space, this evolution is given by
\begin{equation}
\label{unrestricted-eom}
\ddot {\tilde\varphi}_{\k} + 3 H \dot {\tilde\varphi}_{\k}
+ a^{-2} k^2 \tilde\varphi_{\k}=0 ,
\end{equation}
which is the equation of motion of a massless scalar field. In exact de Sitter, the loop contraction is therefore
\be
\label{integral-f} 
\sigma_{\rm tot}^2 (t') =
\frac{1}{2 \pi^2}
\int_0^{\infty} \frac{dk}{k}\, k^3 |f(k,t')|^2 ,
\ee 
where $f(k,t)  = \frac{H}{\sqrt{2 k^3}} \left(1 + \frac{i k}{aH}\right) e^{- i k/aH}$ is the Bunch--Davies mode function solving \eqref{unrestricted-eom}. The singular behavior of this mode at $k=0$ renders \eqref{integral-f} infrared divergent. Formally, this divergence is accompanied by a time-dependent contribution
\be
\label{ds-breaking}
\sigma_{\rm tot}^2 (t') =
{\rm div} + \frac{H^2}{4\pi^2}\ln a(t') ,
\ee
which is the familiar de Sitter-breaking secular growth of the massless two-point function~\cite{Allen:1985ux, Allen:1987tz}.

A standard prescription consistent with (\ref{ds-breaking}) is to integrate (\ref{integral-f}) by introducing a comoving infrared cutoff $k_{\rm IR}$, thereby replacing ``div'' by a finite constant proportional to $\ln(H/k_{\rm IR})$. In the stochastic formulation, this cutoff is often identified with the observational scale $k_*$. This identification, however, has a decisive consequence: it turns the unrestricted field $\varphi$ back into a restricted one. As a result, the mixed contraction entering (\ref{eq:from-qft}) becomes $\langle\varphi_G(t)\varphi(t')\rangle=\sigma_L^2(t')$, rather than \eqref{mixed-contraction}, while the right-hand side of \eqref{ds-breaking} reduces to $\sigma_L^2(t')$. These two replacements precisely recover the cumulants \eqref{stoch_cum}. Thus, the usual prescription is equivalent to discarding the influence of the super-long-wavelength sector $\varphi_{\rm IR}$ on the finite observable sector. It also identifies the infrared scale responsible for de Sitter breaking with the observer-dependent scale $k_*$. Finally, it leaves unresolved the singular limit $k_{\rm IR}\to 0$.

However, introducing a comoving infrared cutoff without the corresponding counterterms is misleading~\cite{Palma:2023idj,Huenupi:2024ksc}. Once the field is endowed with a self-interacting potential $\mathcal V(\varphi)$ bounded from below, the theory develops a finite correlation length $L_\varphi$. Beyond this physical scale, distinct spatial patches become statistically uncorrelated. The scale $L_\varphi$ therefore marks the onset of equilibrium for long-wavelength modes, and with it the loss of any perturbative sensitivity to arbitrarily large distances. Consequently, provided that $k_i>k_*\gg a(t)\,L_\varphi^{-1}$, the connected correlators $\langle\tilde\varphi_{\k_1}\cdots\tilde\varphi_{\k_n}(t)\rangle_c$ cannot depend on the physics of wavelengths larger than $L_\varphi$. It follows that the unrestricted loop integral in \eqref{integral-f}, although formally covering all momenta, cannot be sensitive to the detailed behavior of modes with $k<a(t')L_\varphi^{-1}$. Properly regulated, the loop contraction therefore does not generate secular growth.

To correctly resolve the divergence present in (\ref{integral-f}), in a way compatible with the existence of $L_\varphi$, one can introduce a mass regulator $m_{\rm IR}^2 \ll L_\varphi^{-1}$ in (\ref{unrestricted-eom}). The solution in this case becomes
\be \label{mode-f-massive}
f(k,t) = \frac{i}{2} \sqrt{\frac{\pi}{H}} a^{-3/2} H_\nu^{(1)}\left(\frac{k}{aH}\right),
\ee
where $H_\nu^{(1)}(x)$ is the Hankel function of the first kind with an index $\nu = \sqrt{9/4 - m_{\rm IR}^2 /H^2}$. For $m_{\rm IR}^2>0$ the integral is infrared finite and time independent, yielding the divergent but de Sitter–invariant result:
\be \label{dS-inv-div}
\sigma_{\rm tot}^2\propto m_{\rm IR}^{-2}, \qquad {\rm with } \qquad m_{\rm IR}\to 0 .
\ee

In quasi-de Sitter, the result \eqref{dS-inv-div} is promoted to a slowly varying divergent quantity, with time dependence constrained by \eqref{sigma-slow-roll-suppressed}. To handle the divergent $\sigma_{\rm tot}^2(t')$, one must recall that the coefficients $\lambda_n(t')$ appearing in the bare potential \eqref{Taylor-V} are not physical observables. The relevant quantities are instead the dressed couplings $\bar\lambda_n$, obtained after renormalizing the theory. As shown in~\cite{Huenupi:2024ksc,Huenupi:2024ztu}, the bare potential $\mathcal V$ can be renormalized systematically by splitting $\lambda_n = \bar \lambda_n + \lambda_n^{\rm c.t.}$ and tuning the counterterms $\lambda_n^{\rm c.t.}$ to cancel the divergent contributions generated by $\sigma_{\rm tot}^2$. Order by order, this procedure yields finite dressed couplings
\be
\label{lambda-bar}
\bar\lambda_n (t')
\equiv
\sum_{\ell=0}^{\infty}
\frac{\lambda_{n+2\ell} (t')}{\ell!}
\left(\frac{\sigma_{\rm tot}^2 (t')}{2}\right)^{\ell} .
\ee
These dressed couplings are precisely the Taylor coefficients of the potential
\be
\label{WS-pot}
\bar{\mathcal V}(\varphi , t)
\equiv
\exp\!\left[
\frac{1}{2}\sigma_{\rm tot}^2 (t) \partial_\varphi^2
\right]\mathcal V(\varphi, t),
\ee
namely the Weierstrass transform of $\mathcal V$. The resulting cumulants are finite functions of time
\bea
\label{eq:from-qft-ren}
\Big\langle \varphi_L^n(t) \Big\rangle_c
&=&
-\frac{n}{3} \Big( \sigma_L^2(t) \Big)^{n-1} 
\int_{t_{\rm i}}^{t}\! \frac{\de t'}{H(t')} \bar \lambda_n (t') , 
\eea
which is simply \eqref{eq:from-qft} rewritten after the identification \eqref{lambda-bar}. In exact de Sitter, the integral can be performed explicitly, giving
\be
\label{fi-n-3}
\Big\langle \varphi_L^n(t) \Big\rangle_c
=
-\bar\lambda_n\,
\frac{4\pi^2 n}{3H^4}\,
\sigma_L^{2n}(t).
\ee
This expression differs sharply from \eqref{stoch_cum}. Since $\sigma_{\rm tot}^2$ is time independent in de Sitter, and only slowly varying in quasi-de Sitter, the infrared enhancement of the cumulants in \eqref{fi-n-3} scales as $[\ln a(t)]^n$, in contrast with the loop-enhanced behavior found in \eqref{stoch_cum}.

Notice that (\ref{eq:from-qft-ren}) is equivalent to a tree-level result, whereby one retains only $\ell = 0$ in (\ref{eq:from-langevin}). This is not a coincidence. Single-vertex loop corrections (or tadpoles) vanish in dimensional regularization \cite{Lee:2023jby,Creminelli:2024cge} as well as under normal ordering, leading to physical quantities that are indistinguishable from their tree-level counterparts.


\vspace{3pt}
\noindent\emph{\bf Discussion.--} Our main claim is that the stochastic description of the observable field $\varphi_L$ is incomplete unless the super-long-wavelength sector $\varphi_{\rm IR}$ is treated consistently. The comoving scale $k_\ast$ specifies the range of \emph{external} momenta over which the PDF is defined; it cannot simultaneously serve as an infrared cutoff for \emph{internal} loop momenta. In other words, the observable window is an extrinsic choice, whereas the loop variance is an intrinsic quantity of the quantum theory. Treating the former as a regulator amounts to assuming that coarse graining commutes with the nonlinear potential, thereby discarding the effect of $\varphi_{\rm IR}$.

Restoring this sector leads to a modified Fokker--Planck equation. For instance, in exact de Sitter, where the cumulants are given by \eqref{fi-n-3}, this modification can be written explicitly. Reconstructing the probability density through the Edgeworth expansion \eqref{pdf-qft} and differentiating in time, one obtains~\footnote{In earlier versions of this Letter, we wrote this
equation with the same terms organized in a different way; see also~\cite{Palma:2023idj}. We prefer the present form because it keeps the
potential entirely in the drift.}
\begin{equation} \label{FP}
\dot\rho  = \frac{H^3}{8\pi^2} \rho''   + \frac{1}{3 H} \left[\rho  {\mathcal V}_{\rm eff}\! '(\varphi , t) \right] '  ,
\end{equation}
where $\mathcal V_{\rm eff}(\varphi,t)$ is defined as
\be
\label{V_pdf-def}
\mathcal V_{\rm eff}(\varphi,t)
\equiv  
\exp\!\left[
-\frac{1}{2}\sigma_L^2(t)\partial_\varphi^2
\right] \varphi \partial_{\varphi}\bar{\mathcal V}(\varphi ) .
\ee
Equation~\eqref{FP} has the same differential form as the standard result \eqref{Fokker-Planck}, but differs from it through the explicit dependence on $\sigma_L^2(t)$ entering the drift via \eqref{V_pdf-def}. The derivation can be repeated in quasi-de Sitter, leading to an analogous equation in which the factor $H^3/8\pi^2$ in the diffusion term is replaced by $\frac{1}{2}\frac{\de}{\de t}\sigma_L^2$, while the effective potential takes a slightly more involved form.

This local-in-time equation suggests a corresponding Langevin representation. Once the filtering operation has integrated out both $\varphi_S$ and $\varphi_{\rm IR}$ from the unrestricted field, the standard Langevin equation
\eqref{L-eq-IR} should be replaced by
\be
\label{L-eq-IR-new}
\dot\varphi_L
+\frac{1}{3H} \mathcal V_{\rm eff}'(\varphi , t) 
=H \,\widehat\xi(t),
\ee
with $\mathcal V_{\rm eff}$ given by \eqref{V_pdf-def}. This form, even though obtained perturbatively, may also contain information about the approach to the nonperturbative regime. As argued above, there must exist a correlation length $L_{\varphi}$ such that the perturbative description remains valid as long as $k_*/a(t)>L_{\varphi}^{-1}$. Modes with wavelengths longer than $L_{\varphi}$ are instead governed by nonlinear effects and are expected to settle in equilibrium. It is therefore plausible that, once $k_*/a(t)\sim L_{\varphi}^{-1}$, the potential $\mathcal V_{\rm eff}(\varphi)$ becomes dominated by these modes and ceases to depend explicitly on time. The resulting Fokker--Planck equation would then be autonomous, analogous to \eqref{Fokker-Planck}, but with a dressed potential $\mathcal V_{\rm eff}(\varphi)$ encoding nonperturbative effects. Such an equation would admit the equilibrium solution $\rho_{\rm eq}\propto \exp\left[- \frac{8\pi^2}{3H^4}\mathcal V_{\rm eff}(\varphi)\right]$. This possibility remains speculative, and calls for a dedicated treatment of the $\varphi_{\rm IR}$ sector beyond perturbation theory.

More broadly, our work contributes to a growing effort to delineate the domain of validity of the stochastic formalism. Recent work has emphasized that the nonlinear theory may retain memory of the coarse-graining history, leading to an intrinsically non-Markovian stochastic dynamics \cite{Cruces:2021iwq,Figueroa:2021zah,Cruces:2026yvs}. The effect identified here is complementary: even when a local-in-time description is adopted, modes outside the observable sector reshape the drift through the dressed potential $\mathcal V_{\rm eff}$. A complete formulation should determine how memory effects and drift renormalization combine, and whether they admit a controlled description beyond perturbation theory.

\bigskip
\begin{acknowledgments}

\vspace{3pt}
\noindent\emph{\bf Acknowledgments.--} We wish to thank Sebasti\'an C\'espedes, Xingang Chen, Francisco Colipi, Ian Crosby, Jacopo Fumagalli, Jinn-Ouk Gong, Javier Huenupi, Ellie Hughes, Gabriel Mar\'in, Enrico Pajer, Lucas Pinol, S\'ebastien Renaux-Petel, Bruno Scheihing Hitschfeld, Rodrigo Soto, and Vicharit Yingcharoenrat for useful discussions and comments. We would also like to thank the organizers of the Workshop ``Revisiting cosmological non-linearities in the era of precision surveys" YITP-T-23-03 for inviting us to present this work. GAP acknowledges support from the Fondecyt Regular project 1210876 (ANID). SS is supported by Thailand NSRF via PMU-B [grant number B37G660013].

\end{acknowledgments}


\newpage
\noindent\emph{\bf Supplemental material.--} In the main text, we have shown that infrared modes affect the dynamics of observable degrees of freedom, as encoded in the statistics of $\varphi_L$. The same result can be reached purely within an SPT framework.

First, we have seen that including infrared modes alters the standard Langevin equation~\eqref{L-eq}, leading to~\eqref{L-eq-IR} and, upon integration, to~\eqref{QFT-varphi_L}. To see this from a quantum field theory perspective, consider the full solution
\be
\varphi(\x,t)
=
U(t)\,\varphi_I(\x,t)\,U^\dag(t),
\ee
where $\varphi_I(\x,t)$ is the interaction-picture field satisfying
\be \label{eom-mir}
\ddot\varphi_I+3H\dot\varphi_I-a^{-2}\nabla^2\varphi_I-a^{-2}m_{\rm IR}^2\varphi_I=0,
\ee
with a small positive mass $m_{\rm IR}^2$ introduced as an infrared regulator. The unitary evolution operator is
\be
U(t)
=
T\exp\!\left[
-i\int_{-\infty}^{t}\!\de t'\,
a(t')^3\int_{\x'}\mathcal V\!\left(\varphi_I(\x',t')\right)
\right],
\ee
with $T$ the time-ordering operator.
To first order in the interaction, one finds 
$\varphi(\x,t) \simeq
\varphi_I(\x,t)
+i\int^{t}\!\de t'\!\int_{\x'}
a(t')^3[\varphi_I(\x,t),\varphi_I(\x',t')]
\,\mathcal V'\!\left(\varphi_I(\x',t')\right)$.
Applying the filter $\widehat W$ to restrict the range of observable scales as in Eq.~\eqref{window-2}, we obtain
\be
\label{QFT-varphi_L-2}
\varphi_L(t)
\simeq
\varphi_G(t)
-\frac{1}{3H}\int_{t_{\rm i}}^{t}\!\de t'\,
\widehat W\!\left\{\mathcal V'\!\left[\varphi_I(t')\right]\right\},
\ee
which is equivalent to~\eqref{QFT-varphi_L}, with $\varphi_G=\widehat W\{\varphi_I\}$. This derivation makes explicit the fact that $\varphi_{\rm IR}$ cannot be neglected.

Second, we have shown that the cumulants~\eqref{fi-n-3} follow from Eq.~\eqref{QFT-varphi_L}. These same cumulants can also be obtained directly from standard in-in computations of correlation functions. Recall that the $n$-th cumulant is related to the connected momentum-space correlator $\langle\tilde\varphi_{\k_1\cdots\k_n}(t)\rangle_c$ through
\be
\label{cumulant-n-point}
\Big\langle \varphi^n(t) \Big\rangle_c
\!\! =\!\!
\int_{\k_1}\!W(k_1)\cdots \!\int_{\k_n}\!\!W(k_n)\,
\Big\langle
\tilde\varphi_{\k_1\cdots\k_n}(t)
\Big\rangle_c ,
\ee
with $W(k)$ defined in~\eqref{window-2}. The connected correlator $\langle\tilde\varphi_{\k_1\cdots\k_n}(t)\rangle_c$ can be computed using the action~\eqref{action} and SPT diagrammatic rules \cite{Weinberg:2005vy,Maldacena:2002vr,Chen:2017ryl}. Assuming $m_{\rm IR} \ll k_*(t_{\rm i})/a$, in which case external propagators are described by massless fields, the result is~\cite{Palma:2017lww,Chen:2018uul,Chen:2018brw}
\bea
\label{D-n-point-single-vertex}
\Big\langle
\tilde\varphi_{\k_1\cdots\k_n}(t)
\Big\rangle_c
&=&
(2\pi)^3\delta^{(3)}(\boldsymbol{K})
\frac{2}{3H^4}\frac{H^{2n}}{2^n}
\frac{k_1^3+\cdots+k_n^3}{k_1^3\cdots k_n^3}
\nn\\
&& \!\!\!\!\!\!\!\!\!\!\!\!\!\!\! \times
\ln\!\left(\frac{K}{aH}\right)
\sum_{\ell=0}^{\infty}
\frac{\lambda_{n+2\ell}}{\ell!}
\left(
\frac{1}{2}
\langle\varphi_I\varphi_I\rangle
\right)^{\ell},
\eea
where $\boldsymbol{K}=\k_1+\cdots+\k_n$ and $K=|\boldsymbol{K}|$. The sum over $\ell$ accounts for daisy-loop corrections dressing the single interaction vertex.

The key feature of~\eqref{D-n-point-single-vertex} is that the loop contributions are not restricted by the window function. The latter enters only in~\eqref{cumulant-n-point}, where it limits the external momenta accessible to the observer. By contrast, the correlator $\langle\tilde\varphi_{\k_1\cdots\k_n}(t)\rangle_c$ is an observer-independent quantity intrinsic to the quantum theory, and must therefore respect de Sitter invariance~\cite{Guilleux:2015pma,Palma:2025oux}. Consequently, $\langle\varphi_I\varphi_I\rangle=\sigma_{\rm tot}^2$. Substituting~\eqref{D-n-point-single-vertex} into Eq.~\eqref{cumulant-n-point}, one recovers precisely the cumulants~\eqref{fi-n-3}, in full agreement with the stochastic analysis presented in the text~\cite{Palma:2023idj}.


\begin{thebibliography}{99}

\bibitem{Linde:1982uu}
A.~D.~Linde,
``Scalar Field Fluctuations in Expanding Universe and the New Inflationary Universe Scenario,''
Phys. Lett. B \textbf{116}, 335-339 (1982)

\bibitem{Starobinsky:1982ee}
A.~A.~Starobinsky,
``Dynamics of Phase Transition in the New Inflationary Universe Scenario and Generation of Perturbations,''
Phys. Lett. B \textbf{117}, 175-178 (1982)

\bibitem{Vilenkin:1982wt}
A.~Vilenkin and L.~H.~Ford,
``Gravitational Effects upon Cosmological Phase Transitions,''
Phys. Rev. D \textbf{26}, 1231 (1982)


\bibitem{Ford:1984hs}
L.~H.~Ford,
``Quantum Instability of De Sitter Space-time,''
Phys. Rev. D \textbf{31}, 710 (1985)

\bibitem{Antoniadis:1985pj}
I.~Antoniadis, J.~Iliopoulos and T.~N.~Tomaras,
``Quantum Instability of De Sitter Space,''
Phys. Rev. Lett. \textbf{56}, 1319 (1986)

\bibitem{Weinberg:2005vy}
S.~Weinberg,
``Quantum contributions to cosmological correlations,''
Phys. Rev. D \textbf{72}, 043514 (2005)
[arXiv:hep-th/0506236 [hep-th]].

\bibitem{Weinberg:2006ac}
S.~Weinberg,
``Quantum contributions to cosmological correlations. II. Can these corrections become large?,''
Phys. Rev. D \textbf{74}, 023508 (2006)
[arXiv:hep-th/0605244 [hep-th]].

\bibitem{Burgess:2009bs}
C.~P.~Burgess, L.~Leblond, R.~Holman and S.~Shandera,
``Super-Hubble de Sitter Fluctuations and the Dynamical RG,''
JCAP \textbf{03}, 033 (2010)
[arXiv:0912.1608 [hep-th]].

\bibitem{Seery:2010kh}
D.~Seery,
``Infrared effects in inflationary correlation functions,''
Class. Quant. Grav. \textbf{27}, 124005 (2010)
[arXiv:1005.1649 [astro-ph.CO]].

\bibitem{Polyakov:2012uc}
A.~M.~Polyakov,
``Infrared instability of the de Sitter space,''
[arXiv:1209.4135 [hep-th]].

\bibitem{Hu:2018nxy}
B.~L.~Hu,
``Infrared Behavior of Quantum Fields in Inflationary Cosmology -- Issues and Approaches: an overview,''
[arXiv:1812.11851 [gr-qc]].

\bibitem{Kitamoto:2018dek}
H.~Kitamoto,
``Infrared resummation for derivative interactions in de Sitter space,''
Phys. Rev. D \textbf{100}, no.2, 025020 (2019)
[arXiv:1811.01830 [hep-th]].

\bibitem{Starobinsky:1986fx}
A.~A.~Starobinsky,
``Stochastic de Sitter (inflationary) stage in the early universe,''
Lect. Notes Phys. \textbf{246}, 107-126 (1986)

\bibitem{Starobinsky:1994bd}
A.~A.~Starobinsky and J.~Yokoyama,
``Equilibrium state of a selfinteracting scalar field in the De Sitter background,''
Phys. Rev. D \textbf{50}, 6357-6368 (1994)
[arXiv:astro-ph/9407016 [astro-ph]].


\bibitem{Onemli:2002hr}
V.~K.~Onemli and R.~P.~Woodard,
``Superacceleration from massless, minimally coupled phi**4,''
Class. Quant. Grav. \textbf{19}, 4607 (2002)
[arXiv:gr-qc/0204065 [gr-qc]].

\bibitem{Miao:2005am}
S.~P.~Miao and R.~P.~Woodard,
``The Fermion self-energy during inflation,''
Class. Quant. Grav. \textbf{23}, 1721-1762 (2006)
[arXiv:gr-qc/0511140 [gr-qc]].



\bibitem{Baumgart:2019clc}
M.~Baumgart and R.~Sundrum,
``De Sitter Diagrammar and the Resummation of Time,''
JHEP \textbf{07}, 119 (2020)
[arXiv:1912.09502 [hep-th]].

\bibitem{Palma:2023idj}
G.~A.~Palma and S.~Sypsas,
``Non-Gaussian statistics of de Sitter spectators: a perturbative derivation of stochastic dynamics,''
JHEP \textbf{12}, 170 (2025)
[arXiv:2309.16474 [hep-th]].


\bibitem{Tsamis:2005hd}
N.~C.~Tsamis and R.~P.~Woodard,
``Stochastic quantum gravitational inflation,''
Nucl. Phys. B \textbf{724}, 295-328 (2005)
[arXiv:gr-qc/0505115 [gr-qc]].


\bibitem{Guilleux:2015pma}
M.~Guilleux and J.~Serreau,
``Quantum scalar fields in de Sitter space from the nonperturbative renormalization group,''
Phys. Rev. D \textbf{92}, no.8, 084010 (2015)
[arXiv:1506.06183 [hep-th]].


\bibitem{Finelli:2008zg}
F.~Finelli, G.~Marozzi, A.~A.~Starobinsky, G.~P.~Vacca and G.~Venturi,
``Generation of fluctuations during inflation: Comparison of stochastic and field-theoretic approaches,''
Phys. Rev. D \textbf{79}, 044007 (2009)
[arXiv:0808.1786 [hep-th]].

\bibitem{Riotto:2011sf}
A.~Riotto and M.~S.~Sloth,
``The probability equation for the cosmological comoving curvature perturbation,''
JCAP \textbf{10}, 003 (2011)
[arXiv:1103.5876 [astro-ph.CO]].

\bibitem{Garbrecht:2013coa}
B.~Garbrecht, G.~Rigopoulos and Y.~Zhu,
``Infrared correlations in de Sitter space: Field theoretic versus stochastic approach,''
Phys. Rev. D \textbf{89}, 063506 (2014)
[arXiv:1310.0367 [hep-th]].

\bibitem{Burgess:2014eoa}
C.~P.~Burgess, R.~Holman, G.~Tasinato and M.~Williams,
``EFT Beyond the Horizon: Stochastic Inflation and How Primordial Quantum Fluctuations Go Classical,''
JHEP \textbf{03}, 090 (2015)
[arXiv:1408.5002 [hep-th]].

\bibitem{Garbrecht:2014dca}
B.~Garbrecht, F.~Gautier, G.~Rigopoulos and Y.~Zhu,
``Feynman Diagrams for Stochastic Inflation and Quantum Field Theory in de Sitter Space,''
Phys. Rev. D \textbf{91}, 063520 (2015)
[arXiv:1412.4893 [hep-th]].

\bibitem{Vennin:2015hra}
V.~Vennin and A.~A.~Starobinsky,
``Correlation Functions in Stochastic Inflation,''
Eur. Phys. J. C \textbf{75}, 413 (2015)
[arXiv:1506.04732 [hep-th]].

\bibitem{Burgess:2015ajz}
C.~P.~Burgess, R.~Holman and G.~Tasinato,
``Open EFTs, IR effects \& late-time resummations: systematic corrections in stochastic inflation,''
JHEP \textbf{01}, 153 (2016)
[arXiv:1512.00169 [gr-qc]].

\bibitem{Moss:2016uix}
I.~Moss and G.~Rigopoulos,
``Effective long wavelength scalar dynamics in de Sitter,''
JCAP \textbf{05}, 009 (2017)
[arXiv:1611.07589 [gr-qc]].

\bibitem{Collins:2017haz}
H.~Collins, R.~Holman and T.~Vardanyan,
``The quantum Fokker-Planck equation of stochastic inflation,''
JHEP \textbf{11}, 065 (2017)
[arXiv:1706.07805 [hep-th]].

\bibitem{Tokuda:2017fdh}
J.~Tokuda and T.~Tanaka,
``Statistical nature of infrared dynamics on de Sitter background,''
JCAP \textbf{02}, 014 (2018)
[arXiv:1708.01734 [gr-qc]].

\bibitem{Gorbenko:2019rza}
V.~Gorbenko and L.~Senatore,
``$\lambda \phi^4$ in dS,''
[arXiv:1911.00022 [hep-th]].

\bibitem{Mirbabayi:2019qtx}
M.~Mirbabayi,
``Infrared dynamics of a light scalar field in de Sitter,''
JCAP \textbf{12}, 006 (2020)
[arXiv:1911.00564 [hep-th]].

\bibitem{Mirbabayi:2020vyt}
M.~Mirbabayi,
``Markovian dynamics in de Sitter,''
JCAP \textbf{09}, 038 (2021)
[arXiv:2010.06604 [hep-th]].

\bibitem{Honda:2023unh}
M.~Honda, R.~Jinno, L.~Pinol and K.~Tokeshi,
``Borel resummation of secular divergences in stochastic inflation,''
JHEP \textbf{08}, 060 (2023)
[arXiv:2304.02592 [hep-th]].

\bibitem{PerreaultLevasseur:2013kfq}
L.~Perreault Levasseur,
``Lagrangian formulation of stochastic inflation: Langevin equations, one-loop corrections and a proposed recursive approach,''
Phys. Rev. D \textbf{88} (2013) no.8, 083537
[arXiv:1304.6408 [hep-th]].

\bibitem{Cohen:2020php}
T.~Cohen and D.~Green,
``Soft de Sitter Effective Theory,''
JHEP \textbf{12}, 041 (2020)
[arXiv:2007.03693 [hep-th]].

\bibitem{Cohen:2021fzf}
T.~Cohen, D.~Green, A.~Premkumar and A.~Ridgway,
``Stochastic Inflation at NNLO,''
JHEP \textbf{09}, 159 (2021)
[arXiv:2106.09728 [hep-th]].

\bibitem{Green:2022ovz}
D.~Green,
``EFT for de Sitter Space,''
[arXiv:2210.05820 [hep-th]].

\bibitem{Cohen:2022clv}
T.~Cohen, D.~Green and A.~Premkumar,
``Large deviations in the early Universe,''
Phys. Rev. D \textbf{107}, no.8, 083501 (2023)
[arXiv:2212.02535 [hep-th]].

\bibitem{Ezquiaga:2019ftu}
J.~M.~Ezquiaga, J.~Garc\'\i{}a-Bellido and V.~Vennin,
``The exponential tail of inflationary fluctuations: consequences for primordial black holes,''
JCAP \textbf{03}, 029 (2020)
[arXiv:1912.05399 [astro-ph.CO]].

\bibitem{Vennin:2020kng}
V.~Vennin,
``Stochastic inflation and primordial black holes,''
[arXiv:2009.08715 [astro-ph.CO]].

\bibitem{Hooshangi:2021ubn}
S.~Hooshangi, M.~H.~Namjoo and M.~Noorbala,
``Rare events are nonperturbative: Primordial black holes from heavy-tailed distributions,''
Phys. Lett. B \textbf{834}, 137400 (2022)
[arXiv:2112.04520 [astro-ph.CO]].



\bibitem{Achucarro:2021pdh}
A.~Achucarro, S.~Cespedes, A.~C.~Davis and G.~A.~Palma,
``The hand-made tail: non-perturbative tails from multifield inflation,''
JHEP \textbf{05}, 052 (2022)
[arXiv:2112.14712 [hep-th]].

\bibitem{Cai:2022erk}
Y.~F.~Cai, X.~H.~Ma, M.~Sasaki, D.~G.~Wang and Z.~Zhou,
``Highly non-Gaussian tails and primordial black holes from single-field inflation,''
JCAP \textbf{12}, 034 (2022)
[arXiv:2207.11910 [astro-ph.CO]].

\bibitem{Pi:2022ysn}
S.~Pi and M.~Sasaki,
``Logarithmic Duality of the Curvature Perturbation,''
Phys. Rev. Lett. \textbf{131}, no.1, 011002 (2023)
[arXiv:2211.13932 [astro-ph.CO]].

\bibitem{Hooshangi:2023kss}
S.~Hooshangi, M.~H.~Namjoo and M.~Noorbala,
``Tail diversity from inflation,''
[arXiv:2305.19257 [astro-ph.CO]].

\bibitem{Cruces:2018cvq}
D.~Cruces, C.~Germani and T.~Prokopec,
``Failure of the stochastic approach to inflation beyond slow-roll,''
JCAP \textbf{03}, 048 (2019)
[arXiv:1807.09057 [gr-qc]].


\bibitem{Cruces:2022imf}
D.~Cruces,
``Review on Stochastic Approach to Inflation,''
Universe \textbf{8}, no.6, 334 (2022)
[arXiv:2203.13852 [gr-qc]].

\bibitem{Figueroa:2021zah}
D.~G.~Figueroa, S.~Raatikainen, S.~Rasanen and E.~Tomberg,
``Implications of stochastic effects for primordial black hole production in ultra-slow-roll inflation,''
JCAP \textbf{05}, no.05, 027 (2022)
[arXiv:2111.07437 [astro-ph.CO]].

\bibitem{Cruces:2021iwq}
D.~Cruces and C.~Germani,
``Stochastic inflation at all order in slow-roll parameters: Foundations,''
Phys. Rev. D \textbf{105}, no.2, 023533 (2022)
[arXiv:2107.12735 [gr-qc]].


\bibitem{Launay:2024qsm}
Y.~L.~Launay, G.~I.~Rigopoulos and E.~P.~S.~Shellard,
``Stochastic inflation in general relativity,''
Phys. Rev. D \textbf{109}, no.12, 123523 (2024)
[arXiv:2401.08530 [gr-qc]].

\bibitem{Pinol:2018euk}
L.~Pinol, S.~Renaux-Petel and Y.~Tada,
``Inflationary stochastic anomalies,''
Class. Quant. Grav. \textbf{36}, no.7, 07LT01 (2019)
[arXiv:1806.10126 [gr-qc]].


\bibitem{Chen:2017ryl}
X.~Chen, Y.~Wang and Z.~Z.~Xianyu,
``Schwinger-Keldysh Diagrammatics for Primordial Perturbations,''
JCAP \textbf{12}, 006 (2017)
[arXiv:1703.10166 [hep-th]].

\bibitem{Palma:2017lww}
G.~A.~Palma and W.~Riquelme,
``Axion excursions of the landscape during inflation,''
Phys. Rev. D \textbf{96}, no.2, 023530 (2017)
[arXiv:1701.07918 [hep-th]].

\bibitem{Chen:2018uul}
X.~Chen, G.~A.~Palma, W.~Riquelme, B.~Scheihing Hitschfeld and S.~Sypsas,
``Landscape tomography through primordial non-Gaussianity,''
Phys. Rev. D \textbf{98}, no.8, 083528 (2018)
[arXiv:1804.07315 [hep-th]].

\bibitem{Chen:2018brw}
X.~Chen, G.~A.~Palma, B.~Scheihing Hitschfeld and S.~Sypsas,
``Reconstructing the Inflationary Landscape with Cosmological Data,''
Phys. Rev. Lett. \textbf{121}, no.16, 161302 (2018)
[arXiv:1806.05202 [astro-ph.CO]].


\bibitem{Palma:2025oux}
G.~A.~Palma, S.~Sypsas and D.~Tapia,
``Confronting infrared divergences in de Sitter: loops, logarithms and the stochastic formalism,''
[arXiv:2507.21310 [hep-th]].

\bibitem{Allen:1985ux}
B.~Allen,
``Vacuum States in de Sitter Space,''
Phys. Rev. D \textbf{32}, 3136 (1985)
doi:10.1103/PhysRevD.32.3136

\bibitem{Allen:1987tz}
B.~Allen and A.~Folacci,
``The Massless Minimally Coupled Scalar Field in De Sitter Space,''
Phys. Rev. D \textbf{35}, 3771 (1987)
doi:10.1103/PhysRevD.35.3771

\bibitem{Huenupi:2024ksc}
J.~Huenupi, E.~Hughes, G.~A.~Palma and S.~Sypsas,
``Regularizing infrared divergences in de Sitter spacetime: Loops, dimensional regularization, and cutoffs,''
Phys. Rev. D \textbf{110}, no.12, 123536 (2024)
[arXiv:2406.07610 [hep-th]].

\bibitem{Huenupi:2024ztu}
J.~Huenupi, E.~Hughes, G.~A.~Palma and S.~Sypsas,
``Note on loop resummation in de Sitter spacetime with the wave function of the universe approach,''
Phys. Rev. D \textbf{112}, no.4, 043543 (2025)
[arXiv:2412.01891 [hep-th]].

\bibitem{Lee:2023jby}
M.~H.~G.~Lee, C.~McCulloch and E.~Pajer,
``Leading loops in cosmological correlators,''
JHEP \textbf{11}, 038 (2023)
[arXiv:2305.11228 [hep-th]].

\bibitem{Creminelli:2024cge}
P.~Creminelli, S.~Renaux-Petel, G.~Tambalo and V.~Yingcharoenrat,
``Non-perturbative Wavefunction of the Universe in Inflation with (Resonant) Features,''
[arXiv:2401.10212 [hep-th]].

\bibitem{Cruces:2026yvs}
D.~Cruces and T.~Kuroda,
``A consistent formulation of stochastic inflation I: Non-Markovian effects and issues beyond linear perturbations,''
[arXiv:2605.00476 [astro-ph.CO]].


\end{thebibliography}
\end{document}